\documentclass[aps,twocolumn,pre,showpacs,floatfix]{revtex4}
\usepackage{graphicx}
\newcommand{\be}{\begin{equation}}\newcommand{\bea}{\begin{eqnarray}}
\newcommand{\ee}{\end{equation}}\newcommand{\eea}{\end{eqnarray}}

\begin{document}

\title{
The Effects of Asymmetric Salt and a Cylindrical Macroion on Charge Inversion: 
Electrophoresis by Molecular Dynamics Simulations}

\author{Motohiko Tanaka} \affiliation{ National Institute for Fusion 
Science, Toki 509-5292, Japan} 

\begin{abstract} 
The charge inversion phenomenon is studied by molecular dynamics simulations, 
focusing on size and valence asymmetric salts, and a threshold of surface charge 
density for charge inversion. The charge inversion criteria by the electrophoretic 
mobility and the radial distribution functions of ions coincide except around the 
charge inversion threshold.
The reversed electrophoretic mobility increases with the ratio of coion 
to counterion radii $ a^{-}/a^{+} $, while it decreases with the ratio of 
coion to counterion valences $ Z^{-}/Z^{+} $.  The monovalent salt enhances 
charge inversion of a strongly charged macroion at small salt ionic strength, 
while it reduces reversed mobility otherwise.  A cylindrical macroion is more 
persistent to monovalent salt than a spherical macroion of the same radius and
surface charge density.
\end{abstract}


\date{July 31, 2003}

\maketitle

\section{Introduction}
\label{Sec.1}

The phenomenon of reverting the charge sign of large ions due to other ions 
and salts in water solution was known to physical chemists as charge inversion 
or overscreening for half a century \cite{deJong}.  
More recently, it also attracted a significant attention of physicists 
\cite{Gonza,Elime,Bas,Walker,Bloomf,Kjell,Netz,Gelbart,NguGS,Mainz,Tanaka,Tanaka2,
Tovar,review,Levin,Tovar2}.
Charge inversion is now understood to be the generic phenomenon that occurs 
in strongly correlated charged systems.  It has far reaching consequences in 
biological and chemical worlds.  In particular, it seems to be a decisive 
ingredient in modern gene therapy, facilitating the delivery of genes (negatively 
charged DNA) through cell walls of predominantly negative potentials 
\cite{Kabanov,Safin}.

In our previous papers, we adopted the static \cite{Tanaka} and dynamical  
\cite{Tanaka2} models to study charge inversion of the spherical macroion
by molecular dynamics simulations.  In the former study where a macroion 
was immovable, the static quantities were obtained including the radial 
distribution profiles of counterions and coions moving in the Langevin 
thermostat.  Generally speaking, the radial profiles of charge density and 
integrated charge are good indices of charge inversion.  However, it is 
not so when the charge inversion is around the threshold (a counterexample 
is Fig.\ref{Fig.rdf}(b) of this paper).  
Namely, those ions forming the outskirt of the radial distribution profile 
are not electrostatically bound to the macroion complex and are left behind 
when the complex moves in the solution.  To separate the bound ions from the 
floating ones, the electrophoretic mobility problem in the explicit 
(particle) solvent was studied in the latter.
We showed that a macroion satisfying the charge inversion conditions
drifted with adsorbed counterions and coions toward the external electric 
field, the direction of which agreed with the sign of inverted charge. 

This paper adopts the dynamical (electrophoresis) model and focus on the
following points.  
First, we address the asymmetry effects of the size and valence of counterions 
and coions.  The radii of bare halogen atoms are large compared to those of 
alkaline metals, which become eventually larger than the former after hydration. 
Secondly, we examine the charge inversion threshold in terms of the surface 
charge density and monovalent salt for the macroions of spherical and cylindrical 
shapes \cite{Tanaka3}.  
The adsorption of cylindrical macroions to a charged surface was previously 
studied \cite{Netz,NguGS}.  Instead, we deal with the charge inversion of a 
cylindrical macroion by adsorption of counterions in the presence of monovalent 
salt.  The cylindrical macroion is shown to be more persistent to monovalent 
salt than the spherical macroion of the same radius and surface charge density. 

In the dynamical study \cite{Tanaka2}, the average static quantities including 
the radial distribution profiles of ions and the structure of the electrical 
double layer are not altered in the zeroth order as far as the applied electric 
field $ E $ is much smaller than the electric field produced by the macroion charge, 
$ E \ll Q_{0}/\epsilon R_{0}^{2} $, which is a huge electric field 
$ 10^{6} $V/cm realized by high-power short pulse lasers.  
Here, $ Q_{0} $ and $ R_{0} $ are macroion bare charge and radius, respectively.
We have confirmed this by excellent agreement of the integrated charge profiles 
between the runs with and without the applied electric field, in 3\% accuracy 
for equal-sized coions and trivalent counterions with the parameters of 
Fig.\ref{Fig.rdf1} in Sec.3. 
The net charge of the macroion complex $ Q^{*} $ can be estimated if one makes 
use of the force balance relation, 
\begin{eqnarray}
Q^{*}E - \nu V \approx 0, 
\hspace*{0.2cm} {\rm or} \hspace*{0.2cm}
Q^{*} \approx \nu \mu , 
\end{eqnarray}
where $ \mu = V/E $ is the measured electrophoretic
mobility and $ \nu $ is the solvent friction.  However, one needs to assume the 
friction for this conversion.  The friction that acts on the charged object
in charge-neutral electrolyte solution is an enhanced one compared to the 
Stokesian friction because the velocity decay length is drastically shortened
by screening of hydrodynamic interactions \cite{Ajdari,Viovy}.  
Despite of this uncertainty, the observed mobility agreed qualitatively well 
with the net charge scaling derived by theory \cite{NguGS}.

Other issue of the dynamical study is that the Joule heat is produced by the 
applied electric field, which is transferred to background neutral particles 
through collisions with the ions.  
To drain the heat, we adopt a heat bath at the domain boundary 
that is kept at a constant temperature $ T $.  The thermal bath screens 
hydrodynamic interactions and nullifies the momentum of the system on time
average.  However, it is emphasized that the electric field does not bring in 
momentum into the charge-neutral system, and that hydrodynamic interactions 
are weak compared to the electrostatic binding forces and screened at short 
distances in the electrolyte solution \cite{Ajdari,Viovy}.  
Momentum is not transported beyond this distance from the macroion toward the 
domain boundaries.  
We confirmed this fact by comparing the mobilities of two runs with and without 
the heat bath (before heating became significant), the results of which agreed 
very well within simulation errors (cf. Fig.5 of \cite{Tanaka2}).  Thus, the 
heat bath does not affect our simulations.  

This paper is organized as follows. The simulation method and parameters 
are described in Sec.\ref{Sec.2}. The effects of asymmetric radii and valences 
of coions and counterions on the charge inversion phenomenon are studied
in Sec.\ref{Sec.3}.
The dynamical and static observables, i.e. electrophoretic mobility and
radial distribution functions of ions, are compared for the same runs,
where good correspondences are found except around the charge inversion
threshold.  
In Sec.\ref{Sec.4}, the effects of monovalent salt that exits as the base of 
the Z:1 counterions and coions are examined.  
There, a cylindrical {\it infinite} macroion (occupying full length across 
the domain) that can adsorb geometrically more counterions than a spherical 
macroion of the same radius is introduced.
The cylindrical macroion is found to be more persistent to monovalent salt
than the spherical macroion of the same radius and surface charge density.
Sec.\ref{Sec.5} will be a summary of this paper.

\vspace*{-0.3cm}
\section{Simulation Method and Parameters}
\label{Sec.2}

Simulation method and parameters for the present study are described here.  
We take the system consisting of one macroion, many counterions, coions and 
neutral particles. We solve the Newton equations of motion for each
particle with the Coulombic and Lennard-Jones potential forces 
under a uniform applied electric field $ E $ ($ E > 0 $). 
A large number of neutral particles are used to model the solvent of given 
temperature and to treat the interactions among the finite-size macroion, 
counterions and coions.  

The units of length, charge and mass are, $ a $, $ e $, and $ m $,
respectively.  As is mentioned below, our choice of the temperature 
corresponds to $ a \sim 1.4 $\AA \ in water and $ m \sim 40 $ a.m.u.
A spherical macroion with radius $ R_{0}= 5a $, negative charge $ Q_{0} $
between $ -15e $ and $ -81e $, and mass $ 200m $ is surrounded by 
the $ N^{+} $ number of counterions of a positive charge $ Z^{+}e $ 
and the $ N^{-} $ coions of a negative charge $ -Z^{-}e $. 
The corresponding surface charge density of the macroion is between
$ \sigma_{sp}= 0.048e/a^{2} $ (0.39 C/m$ {}^{2} $) and $ 0.26e/a^{2} $
(2.1 C/m$ {}^{2} $).  A rod-shaped macroion is also used in Sec.\ref{Sec.4}
whose surface charge density is between $ \sigma_{rod}= 0.04e/a^{2} $ 
(0.33 C/m$ {}^{2} $) and $ 0.08e/a^{2} $ (0.66 C/m$ {}^{2} $).
The system is maintained in overall charge neutrality, $ Q_{0} + 
N^{+}Z^{+}e - N^{-}Z^{-}e = 0 $.  
The radii of counterions and coions are $ a^{+} $ and $ a^{-} $, 
respectively, with the counterion radius being fixed at $ a^{+}= a $, 
and the radius of neutral particles is $ a/2 $.  The mass of the coions 
and counterions is $ m $, and that of $ N_{*} $ neutral particles is 
$ m/2 $, where we have in mind the water molecule against 
that of K$ {}^{+} $ or Ca$ {}^{2+} $ ions.  However, the mass is
not involved as far as an equilibrium state is concerned.
Approximately one neutral particle is distributed in every volume
element $ (2.1a)^{3} \approx (3 $\AA$)^{3} $ inside the simulation
domain, excluding the locations already occupied by ions, which 
typically yields 4000 neutral particles for the spherical macroion
of radius $ R_{0}= 5a $.
These particles are placed in a cubic box of size $ L=32a$,
with periodic boundary conditions in all three directions.

Calculation of the Coulomb forces under the periodic boundary
conditions requires the charge sum in the first Brillouin zone
and their infinite number of mirror images (the Ewald sum\cite{Ewald}).
The sum is calculated with the use of the PPPM algorithm
\cite{Eastwood,Deserno}.
We use $ (32)^{3} $ spatial meshes for the calculation of the
reciprocal space contributions to the Coulomb forces, with
the choice of the Ewald parameter $ \alpha \approx 0.262 $ and the
cutoff length $ 10a $.

In addition to the Coulomb forces, the volume exclusion effect among 
particles is introduced through the repulsive Lennard-Jones potential 
$ \phi_{LJ}= 4 \varepsilon [(A/r_{ij} )^{12} - ( A/r_{ij} )^{6}] $ 
for $ r_{ij}= |{\bf r}_{i}-{\bf r}_{j}| \le 2^{1/6}A $, and $ \phi_{LJ}= 
-\varepsilon $ otherwise to exclude the attraction part. 
Here $ {\bf r}_{i}$ is the position vector of the $ i $-th particle, 
and $ A $ is the sum of the radii of two interacting particles. 
We relate $ \varepsilon $ with the temperature by $ \varepsilon = k_{B}T $, 
and choose $ k_{B}T = e^{2}/5 \epsilon a $ (we assume spatially homogeneous 
dielectric constant $ \epsilon $).  The Bjerrum length is thus $ \lambda_{B}= 
e^{2}/\epsilon k_{B}T= 5a $, which implies $ a \approx 1.4 $\AA \ in water.  

The initial states of the runs are prepared by randomly positioning all 
the ions and neutral particles in the simulation domain and giving them
Maxwell-distributed random velocities.  The Newton equations of motion 
are integrated with the use of the leapfrog method, which is equivalent 
to the Verlet algorithm \cite{textMD}. 
The unit of time is $ \tau = a \sqrt{m/ \varepsilon} $, and we choose 
the integration time step $ \Delta t= 0.01 \tau $, where $ \tau \approx
1 ps $ in real environments. 
The simulation runs are executed well beyond the time when the dynamical 
property (the macroion drift speed in this study) has become stationary,
and only the stationary part is used for data analysis. 
Under the electrostatic and Lennard-Jones forces where the random forces
and frictional forces are represented by particle solvent, typical run times 
are between $ 2000 \tau $ and $ 4000 \tau $ (some runs are continued 
further to check the long term variability, but the results fall within 
the error bars). 

As noted earlier, we can adopt the heat bath without side effects for the 
present study with the neutral electrolyte solvent. At the center of the
heat bath the macroion is located at every moment.  The velocities of the 
neutral particles crossing the boundaries are refreshed according to the 
thermal distribution of the given temperature.
Throughout this paper, the same normalization of the mobility $ \mu_{0}= 
v_{0}/(|Q_{00}|/\epsilon R_{0}^{2}) \approx 21 (\mu{\rm m/sec})/({\rm V/cm}) $ 
is used, where $ Q_{00}= -80e $, $ R_{0}= 5a $ and $ v_{0} $ is the thermal
speed of neutral (solvent) particles.

\vspace*{-0.2cm}
\section{The Effects of Asymmetric Coions/Counterions}
\label{Sec.3}

\begin{figure}
\centerline{\scalebox{0.75}{\includegraphics{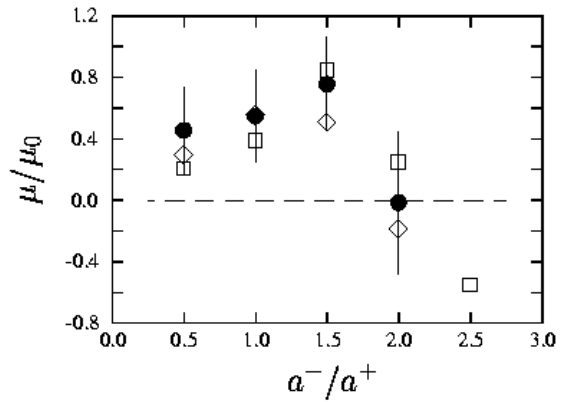}}}

\vspace*{-0.35cm}
\caption{ The electrophoretic mobility of the macroion $ \mu $ is shown against 
the ratio of coion and counterion radii $ a^{-}/a^{+} $. The counterion radius 
is fixed at $ a^{+}= a $, and their valence is $ Z^{+}= $ 2 (diamonds), 
3 (circles), and 5 (squares). The surface charge density of the macroion is 
$ \sigma_{sp} \sim 0.26e/a^{2} $ (2.1 C/m$ {}^{2} $) for the charge $ Q_{0}= -80e $ 
and radius $ R_{0}= 5a $.  Here, the normalization of the mobility is
$ \mu_{0}= v_{0}/ (|Q_{0}|/\epsilon R_{0}^{2}) \approx 21 (\mu{\rm m/sec})/
({\rm V/cm}) $ with $ v_{0} $ the thermal speed of neutral particles.  
The number of monovalent coions is $ N^{-} \approx 60 $ ($ 1.1 $Mol/l). 
The Bjerrum length is $ \lambda_{B}= 5a $.}
\label{Fig.mu_acoi}
\end{figure}

The dependence of the electrophoretic mobility of a macroion on 
the ratio of coion and counterion radii $ a^{-}/a^{+} $ is shown in 
Fig.\ref{Fig.mu_acoi}.  
The coion radius is varied between $ 0.5a $ and $ 2.5a $ for the 
fixed counterion radius $ a^{+}= a $, and the Bjerrum length is
$ \lambda_{B}= e^{2}/\epsilon k_{B}T= 5a $.  The counterion valence 
is $ Z^{+}= $ 2 (diamonds), 3 (circles) and 5 (squares).  The number 
of monovalent coions is $ N^{-} \sim 60 $, which corresponds to the
density $ 1.8 \times 10^{-3}/a^{3} \approx $1.1 Mol/l, and that of 
the counterions is determined by charge neutrality.
The macroion with charge $ Q_{0} = -80e $ and radius $ R_{0}= 5a $ has 
the surface charge density $ \sigma_{sp} \sim 0.26e/a^{2} $ (2.1 C/m$ {}^{2} $). 
The electric field is $ E= 0.1 \varepsilon /ae \ll Q_{0}/\epsilon R_{0}^{2} $ 
which ensures that the electrophoresis occurs in the linear regime
\cite{Tanaka2}.

Figure \ref{Fig.mu_acoi} shows that the mobility is reversed (positive)
and increases almost linearly with the ratio $ a^{-}/a^{+} $.  
The mobility peaks at the intermediate coion radius $ a^{-}/a^{+} \approx 1.5 $, 
beyond which the mobility sharply decreases and flips to non-reversed.  
The degradation of the mobility does not depend on the counterion valence nor 
is indexed to the coion and counterion interaction energy $ Z^{+}Z^{-}e^{2}/
\epsilon(a^{-}+a^{+}) $ which is always larger than thermal energy $ k_{B}T $ 
(between 5 and 10 times for $ Z^{+}= 3 $). 
On the other hand, the volume fraction of the ions becomes significant, 
$ 4\pi(N^{-}(a^{-})^{3} +N^{+}(a^{+})^{3})/3L^{3} \sim 0.07 $ for $ a^{-}= 2a $
and 0.13 for $ a^{-}= 2.5a $.  

The reversed mobility at $ a^{-}= a^{+} $ and $ Z^{+}= 3 $ is 
12 $ (\mu{\rm m/sec})/({\rm V/cm}) $.  
The experiment with a sulphonated latex particle of the surface charge density 
0.115 C/m$ {}^{2} $ obtained the saturated mobility 0.25 $ (\mu{\rm m/sec})/
({\rm V/cm}) $ for 50 mM/l La(NO$ {}_{3}$)$ {}_{3} $ (Fig.6 of Ref.\cite{Tovar2}).  
Although exact comparison is difficult, we note that the theory predicts
\cite{NguGS} that net charge $ Q^{*} $, hence the mobility, scales as 
$ Q^{*}/R_{0}^{2} \sim \sigma (R_{WS}/\lambda_{D}) \sim (Ze \sigma c_{s})^{1/2} $ 
for the weak Debye shielding regime, where $ \lambda_{D} = 
(\epsilon k_{B}T /8\pi c_{s}e^{2})^{1/2} $ with $ c_{s} $ the coion density, 
and $ R_{WS} = (Ze/\pi \sigma_{sp})^{1/2} $ is the Wigner-Seitz cell radius. 
This formula may be applicable because of weak lateral screening on the macroion 
surface due to the absence of coions there.
The mobility of the former is expected to be 37 times that of the latter,
whose agreement is fair. 

\begin{figure}
\centerline{\scalebox{0.75}{\includegraphics{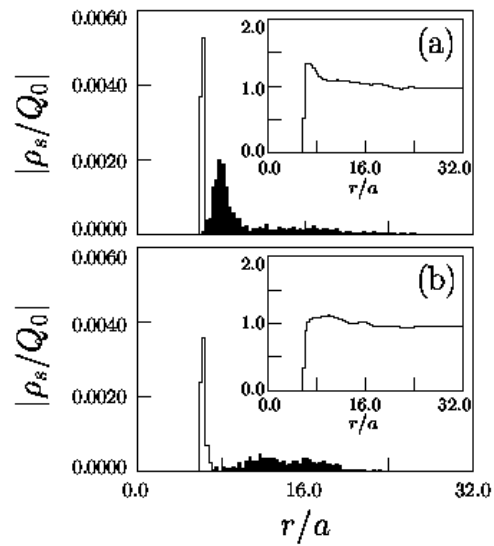}}}
\vspace*{-0.35cm}
\caption{ The radial distribution functions of charge density $ \rho_{s} $ of 
counterions (solid bars) and coions (shaded bars), with the coion radius (a) 
$ a^{-}= a $ and (b) $ a^{-}= 2a $, are shown for the runs in Fig.\ref{Fig.mu_acoi} 
(the coion bars are enlarged by ten times).  The counterion radius is 
$ a^{+}=a $ and valence $ Z^{+}= 3 $.  The inset panels show the integrated charge 
distribution $ Q(r)/|Q_{0}| $ of the corresponding counterion and coion charge 
densities.}
\label{Fig.rdf1}
\end{figure}

The cases with reversed mobility are characterized by the peaked radial distribution 
profile of the counterions localized on the macroion surface and strong 
association of coions to the counterions, as seen in Fig.\ref{Fig.rdf1}(a).
The integrated charge $ Q(r) $ in the inset panel is peaked and larger than the 
neutrality $ |Q_{0}| $.  By contrast, for the case of non-reversed mobility at 
$ a^{-}=2a $ in Fig.\ref{Fig.rdf1}(b), the aggregates made of one counterion 
and a few condensed large coions distribute more homogeneously and apart from
the macroion than for the $ a^{-}/a^{+}= 1 $ case. The radial profile of the 
integrated charge is nearly flat due to less concentrated counterions, thus
the dynamical and static observables of charge inversion are consistent.

The increase in the mobility with coion radius arises from less amount of
charge compensation by large coions due to geometrical avoidance on the 
counterion surface.  This is consistent with the Monte Carlo simulation 
of charge inversion for the finite-size coions \cite{Kjell} and the condensation 
of the $ Z:1 $ ions with the size asymmetry \cite{Fisher}. 
Our observation shows that the degradation of the mobility for large coions is
due to collisions; a part of adsorbed counterions are kicked out of the macroion 
surface whose return is hampered and delayed by large coions surrounding the 
macroion.  
This is contrary to the charge inversion enhancement by the excluded volume 
ordering by large and symmetric coions and counterions \cite{Messina2}.    

\begin{figure}
\centerline{\scalebox{0.70}{\includegraphics{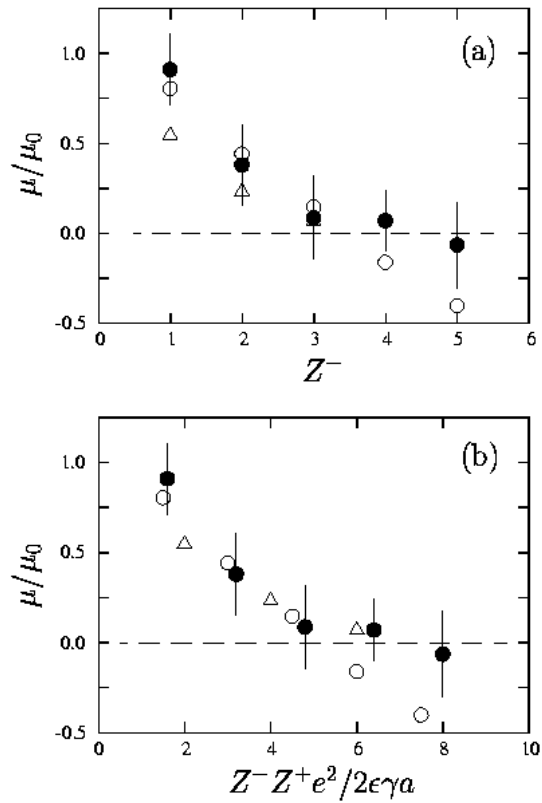}}}

\vspace*{-0.35cm}
\caption{The macroion mobility is shown against (a) the coion valence $ Z^{-} $,
and (b) the modified interaction energy of counterions and coions  
$ Z^{+}Z^{-}e^{2}/ 2\epsilon \gamma a $,
for the fixed counterion valences, $ Z^{+}= 2 $ (triangles), $ Z^{+}= 3 $ 
(open circles), and $ Z^{+}= 4 $ (solid circles). The $ \gamma $ factor is
1.0, 2.0 and 2.5 for $ Z^{+}= $ 2, 3 and 4, respectively.
Coions and counterions are of the same radius, $ a^{-}= a^{+} = a $, and 
the temperature is $ e^2/\epsilon ak_{B}T= 5 $ (the Bjerrum length 
$ \lambda_{B}= 5a $).}
\label{Fig.mu_coval}
\end{figure}

The effect of asymmetric valences between counterions and coions on the 
electrophoretic mobility is shown in Fig.\ref{Fig.mu_coval}.  The coion valence 
$ Z^{-} $ is varied for the fixed counterion valence $ Z^{+} $ which is either 
divalent, trivalent or tetravalent.  The macroion surface charge density is 
$ \sigma_{sp} \sim 0.26e/a^{2} $ (2.1 C/m$ {}^{2} $) for the charge and radius 
$ Q_{0} \sim -80e $ and $ R_{0}= 5a $.
The charge content carried by coions is kept the same such that $ N^{-}= 
300/Z^{-} $, and that of the counterions is determined by charge neutrality. 
The radius of the coions and counterions is equal, $ a^{-}= a^{+}= a $. 

Figure \ref{Fig.mu_coval}(a) shows that the mobility is reversed and largest 
for the monovalent coions, i.e. for the largest asymmetry with any $ Z^{+} $ 
value.  The mobility becomes larger as the valence of the counterions 
increases from divalent, trivalent to tetravalent at $ Z^{-}= 1 $.
As the degree of the asymmetry decreases with the increase in coion valence, 
the magnitude of reversed mobility decreases as $ (Z^{-})^{-1} $ until the 
two valences become nearly symmetric, $ Z^{+} \sim Z^{-} $.  
The mobility remains reversed for the equal valence cases except for 
$ Z^{+} = Z^{-} = 1 $, and finally becomes non-reversed for larger coion 
valences.  
If Fig.\ref{Fig.mu_coval}(a) is replotted for the coion and counterion 
interaction energy modified by a factor $ \gamma $ as in 
Fig.\ref{Fig.mu_coval}(b), three curves of different valences $ Z^{+} $
follow a master curve, where $ \gamma = $ 1, 2 and 2.5 for $ Z^{+}= $
2, 3 and 4, respectively.  The $ \gamma $ factor stands for the reduction of
the interaction energy by enhanced screening possibly due to condensed
coions to the counterions, which is actually observed. 

The charge inversion for divalent counterions and coions is in line with the 
theory result of the HNC-MSA integral equations \cite{Tovar}.  The radial
distribution profile of the integrated charge is peaked $ Q_{peak}/Q_{0} \sim
1.4 $, and its trailing tail oscillates periodically around the neutrality 
as found previously \cite{Tanaka,Messina2}.  
Although the $ Z^{+}=Z^{-} $ case resembles the Debye screening in the sense 
that the ratio of the macroion, counterion and coion valences is $ Q_{0}/Z:1:1 $, 
neither the Debye or nonlinear Poisson-Boltzmann theory is applicable 
due to strong electrostatic interactions and charge inversion takes place. 

\begin{figure}
\centerline{\scalebox{0.75}{\includegraphics{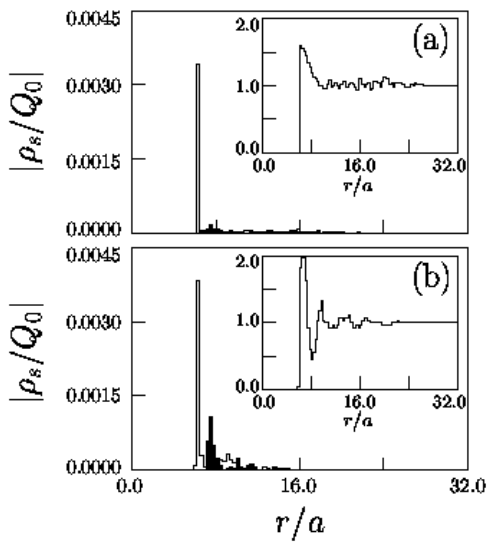}}}
\vspace*{-0.35cm}
\caption{ The radial distribution functions of charge density $ \rho_{s} $ 
of counterions (solid bars) and coions (shaded bars), with the coion valence 
(a) $ Z^{-}= 1 $ and (b) $ Z^{-}= 3 $, are shown for the runs in 
Fig.\ref{Fig.mu_coval}.  The counterion valence is $ Z^{+}= 3 $.  
The inset panels show the integrated charge distribution $ Q(r)/|Q_{0}| $ 
of the corresponding counterion and coion charge densities.}
\label{Fig.rdf}
\end{figure}

The dynamical results of the mobility in Fig.\ref{Fig.mu_coval} are compared 
with the static results in Fig.\ref{Fig.rdf} that depicts the radial 
distribution functions for trivalent counterions and the coions with (a) 
$ Z^{-}= 1 $ and (b) $ Z^{-}= 3 $.  In the panel (a), the counterions are 
strongly adsorbed to the macroion and the radial profile of the integrated charge 
in the inset panel is peaked, $ |Q_{peak}/Q_{0}| \approx 1.6 $.
Both the dynamical and static observables indicate strong charge inversion.  
It is counterintuitive, however, that the maximum peak height of the integrated 
charge in Fig.\ref{Fig.rdf}(b) is not proportional to the small mobility 
measured in Fig.\ref{Fig.mu_coval}.  
The integrated charge profile with a sharp dip after the first peak is
usually associated with small reversed or non-reversed mobilities.
For crudely detecting the charge inversion statically, one has to count the number 
of the ions whose binding potential to the complex is larger than $ k_{B}T $,
which is not trivial in data analysis of the molecular dynamics simulations.

\vspace*{-0.2cm}
\section{Spherical and Cylindrical Macroions under Monovalent Salt}
\label{Sec.4}

We study in this section both the spherical and cylindrical macroions under 
the monovalent salt that exists as the base component to the Z:1 multivalent 
salt.  Here, the trivalent counterions ($ Z= 3 $) are called the Z-ions, 
and other counterions and coions are monovalent.  All these ions have equal 
radius $ a^{-}= a^{+}= a $. 

\begin{figure}
\centerline{\scalebox{0.74}{\includegraphics{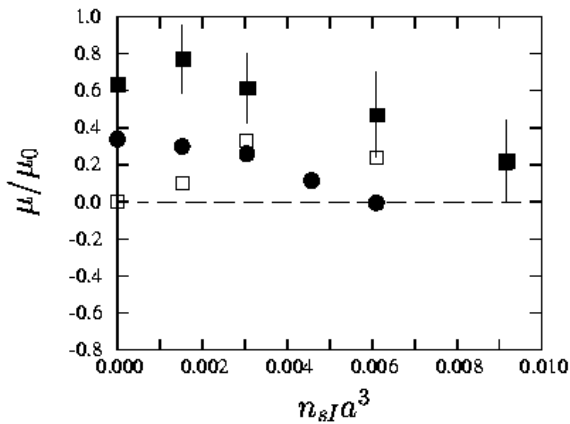}}}

\vspace*{-0.35cm}
\caption{ The electrophoretic mobility of the spherical macroion is shown
against ionic strength of monovalent salt $ n_{sI} $ (1 Mol/l salt 
corresponds to $ 0.0017/a^{3} $).  The surface charge density of the 
macroion is $ \sigma_{sp}= Q_{0}/4 \pi R_{0}^{2} \sim 0.26e/a^{2} $, with 
excess Z-ions (filled squares), and without excess Z-ions (open squares). 
Also, the mobility for the macroion $ \sigma_{sp} \sim 0.080e/a^{2} $ with 
excess Z-ions (filled circles) is shown.  
Here, $ \mu_{0} \approx 21 (\mu{\rm m/sec})/({\rm V/cm}) $.}
\label{Fig.mu_1salt}
\end{figure}

\begin{figure}
\centerline{\scalebox{0.75}{\includegraphics{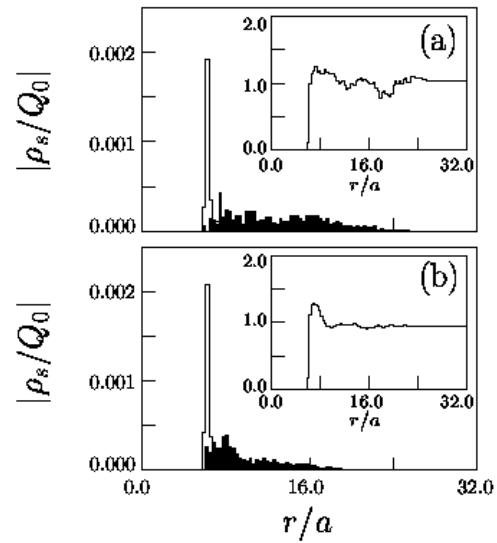}}}

\vspace*{-0.35cm}
\caption{ The radial distribution functions of charge density $ \rho_{s} $ 
of counterions (solid bars) and coions (shaded bars) are shown for 
(a) the spherical macroion in Fig.\ref{Fig.mu_1salt}, 
and (b) the cylindrical macroion in Fig.\ref{Fig.mu_Rsalt}, both with
surface charge density $ \sigma = 0.08e/a^{2} $, radius $ R= 5a $, and 
salt ionic strength $ n_{sI} \sim 0.006/a^{3} $
(the coion bars are enlarged by three times). 
The inset panels show the integrated charge distribution $ Q(r)/|Q_{0}| $ 
of the corresponding counterion and coion charge densities.}
\label{Fig.rdf-sal}
\end{figure}

\begin{figure}
\centerline{\scalebox{0.75}{\includegraphics{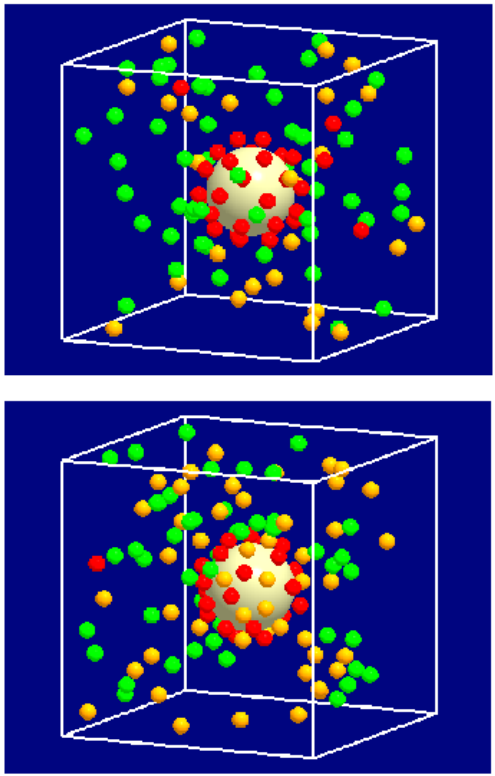}}}

\vspace*{-0.35cm}
\caption{ The bird's-eye views of all ion species are shown for the spherical 
macroion with (top) and without (bottom) the excess Z-ions for the macroion of
surface charge density $ \sigma_{sp} \sim 0.26e/a^{2} $ in Fig.\ref{Fig.mu_1salt}. 
The trivalent and monovalent counterions are drawn in red and yellow, 
respectively, and the monovalent coions in green (nearly 4000 solvent particles
are not drawn). }
\label{Fig.bird2_sal}
\end{figure}

Figure \ref{Fig.mu_1salt} shows the dependence of the mobility of a 
spherical macroion against ionic strength of monovalent salt, 
$ n_{sI}= 2N^{+1}/L^{3} $, where $ N^{+1} $ is the number of monovalent 
counterions (the same number of matching monovalent coions are present).  
The $ N^{+1}= 50 $ salt ions correspond to $ n_{sI} \sim 0.0031/a^{3} $ 
(1.8 Mol/l), and the Debye screening length is $ 1.6 a $.
The surface charge density for the {\it strongly} charged macroion with 
$ Q_{0}=-81e $ (filled and open squares) is $ \sigma_{sp} \sim 0.26e/a^{2} $, 
and that for the {\it weakly} charged macroion with $ Q_{0}=-25e $ (circles)
is $ 0.08e/a^{2} $ (0.65 C/m$ {}^{2} $), both with the radius $ R_{0}= 5a $.  
For these runs, the surface charge is at least several times that of the salt 
ions contained in the surface layer, $ \sigma_{sp} \gg en_{sI} \lambda_{D} $. 
The Guoy-Chapman length is even smaller, $ \lambda_{GC}=
\epsilon k_{B}T/2\pi Ze \sigma_{sp} \sim 0.13a $ for trivalent counterions
and $ \sigma_{sp} \sim 0.26e/a^{2} $.  
The Wigner-Seitz cell radius for the spherical macroion is,
\begin{eqnarray}
R_{WS} = (Ze/\pi \sigma_{sp})^{1/2} = 2R_{0}(Ze/|Q_{0}|)^{1/2}. 
\end{eqnarray}
This yields $ R_{WS} \sim 1.9a $ and $ 3.5a $, respectively, for the 
{\it strongly} and {\it weakly} charged macroions in Fig.\ref{Fig.mu_1salt}.
A series of the runs with spherical macroions of different bare charges 
$ Q_{0} $ reveal that the charge inversion threshold under the zero monovalent 
salt is $ Q_{0} \sim 15e $ for $ R_{0}= 5a $.  This yields $ Q_{0}/R_{0}^{2}
\sim 0.6e/a^{2} $ and the surface charge density at the threshold
\begin{eqnarray}
\sigma_{sp} \sim 0.048 e/a^{2} \ (0.39 {\rm C/m}^{2}).
\end{eqnarray}
This agrees with our previous finding \cite{Tanaka2} that the 
threshold of surface charge density is $ |Q_{0}|/R_{0}^{2} \approx 0.5e/a^{2} $ 
for any combination of $ Q_{0} $ and $ R_{0} $.  The correlation energy of 
the surface Z-ions forming the Wigner-Seitz cell becomes at the threshold
\begin{eqnarray}
Z^{2}e^{2}/2\epsilon R_{WS} \sim 5 k_{B}T.  
\label{eq:ze}
\end{eqnarray}

In Fig.\ref{Fig.mu_1salt}, we see that the addition of small amount of 
monovalent salt enhances the reversed mobility for the strongly charged 
macroion with excess Z-ions $ N_{b}^{+3}=10 $ (filled squares).
The mobility peaks at $ n_{sI} \sim 0.0015e/a^{2} $ (0.83 Mol/l).
However, further addition of monovalent salt whose charge concentration 
exceeds that of the Z-ions in the bulk screens the electric field 
and suppresses the charge inversion. 
By extrapolation, the reversed mobility is expected to terminate at 
$ n_{sI} \sim 0.013/a^{3} $ (7.2 Mol/l).
Even without the excess Z-ions for which $ N^{+3}= |Q_{0}|/eZ $, the 
addition of monovalent salt induces the charge inversion (open squares;
the fifth data point overlaps on the filled square). 
On the other hand, for a weakly charged macroion with excess trivalent Z-ions 
$ N_{b}^{+3}= 10 $, the mobility is reversed at zero salt but decreases 
monotonically with the salt ionic strength. 

A close look at the ion distributions shows that the enhancement of the 
mobility at the {\it small} salt ionic strength in Fig.\ref{Fig.mu_1salt} 
is due to the rearrangement of ions to achieve global energy minimization 
rather than the adjustment of the Z-ions on the macroion surface.  
This is found in the radial distribution functions of ions in 
Fig.\ref{Fig.rdf-sal}(a), 
and visually in the bird's-eye view plot in Fig.\ref{Fig.bird2_sal}. 
For the case with excess Z-ions, almost all the Z-ions are adsorbed to the 
macroion, while monovalent salt ions including counterions are located 
apart from the macroion surface.  
For the case without the excess Z-ions, both trivalent and monovalent 
counterions are adsorbed to the macroion, but the coions are absent in 
the vicinity of the macroion.  Thus, the lateral screening on the macroion 
surface should not be significant.  
The theory of charge inversion \cite{NguGS} predicts enhancement of 
overcharging by monovalent salt. However, the theory heavily relies on 
the screening of surface Z-ion correlations beyond the nearest neighbor
cells, and the predicted overcharging occurs in the limited regime. 

We interpret that the aforementioned rapid decrease in the mobility 
with salt ionic strength for the weakly charged macroion is due to the 
finite radius of the spherical macroion.  Namely, a number of Z-ions 
adsorbed on the small surface of the macroion may not be sufficient 
to maintain electrostatic correlations and overcome thermal fluctuations.

Therefore, the rest of this paper is devoted to the charge inversion 
of a cylindrical macroion.  Here, the macroion is assumed to be a rod
of finite radius $ R_{rod} $ and {\it infinite} length, occupying 
the full length across the domain, and that it lies perpendicularly to 
the applied electric field for the ease of the simulation. 
It is remarked that a polyelectrolyte chain tends to align along the 
electric field due to the polarization effect \cite{Netz2}.  However, 
the orientation of the rod may be of the second importance in the case
of a non-deformable rod macroion since the counterion adsorption by 
the macroion is not much altered as far as the applied electric field 
is weak compared to the local electric field produced by the macroion, 
$ E \ll Q_{rod}/\epsilon R_{rod}L $.  

Before other results are shown, the dependence of the mobility on the length 
of the rod is examined.  Table I summarizes the mobility for the finite
length rod which has the cylinder length $ \ell $ and is capped by two 
semi-spheres of radius $ R $ at each end.  The surface area of the macroion 
is $ S= 2\pi R(2R +\ell) $, and the fixed surface charge density $ \sigma_{rod}
\approx 0.08e/a^{2} $ determines the macroion charge $ Q_{rod} $. 
The $ \ell= 0 $ case is a spherical macroion.
The surface electric fields at the center and tips of the rod are thus
almost the same for different rods. The mass of each rod is varied 
such that the ratio $ Q_{rod}/M_{rod} \approx const. $ although
the rod mass should not be involved in the equilibrium drift speed
and the mobility.  
Table I shows that the mobility slightly increases for a short rod compared
to the spherical macroion but becomes nearly insensitive to the length of 
the rod for $ \ell /R_{rod} \ge 3 $, if the surface charge density is 
maintained and the monovalent salt is not present.  
This rod length may indicate the transition from the 3-D to 2-D regime
of the cylindrical macroion.

\begin{table} 
\caption{The dependence of mobility $ \mu $ on the length $ \ell $ of 
the cylindrical macroion with the radius $ R= 5a $ that lies perpendicularly 
to the applied electric field.  The surface charge density $ \sigma_{rod} 
\approx 0.08e/a^{2} $ gives the macroion charge $ Q_{rod} = \sigma S $ for 
the surface area $ S= 2\pi R(2R +\ell) $. The number of coions is $ N^{-} 
\approx 90 $ and monovalent salt is not present.} 

\vspace*{0.3cm}
\begin{tabular}{crrrrr} \hline \hline
\hspace{0.3cm} $ \ell/a $ \hspace{0.3cm} & 0 \hspace{0.3cm} & 5 \hspace{0.3cm} & 
10 \hspace{0.3cm} & 15 \hspace{0.3cm} & 20 \hspace{0.3cm} \\
\hline
$ S/a^{2}$  & 314 \hspace{0.3cm} & 471 \hspace{0.3cm} & 628 \hspace{0.3cm} & 
 785 \hspace{0.3cm} & 942 \hspace{0.3cm} \\  
$ |Q_{rod}|/e $  &  25 \hspace{0.3cm} & 38 \hspace{0.3cm} & 50 \hspace{0.3cm} & 
  63 \hspace{0.3cm} & 75 \hspace{0.3cm} \\
\hline
$ \mu /\mu_{0} $ &  0.29 \hspace{0.3cm} & 0.35 \hspace{0.3cm} & 0.34 \hspace{0.3cm} 
& 0.31 \hspace{0.3cm} & 0.30 \hspace{0.3cm} \\
\hline \hline
\end{tabular}
\end{table}

\begin{figure}
\centerline{\scalebox{0.74}{\includegraphics{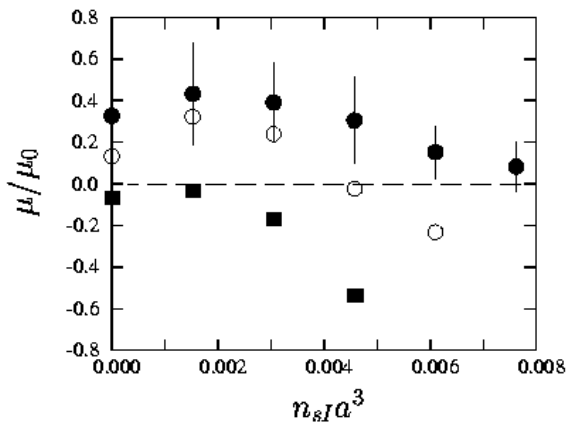}}}

\vspace*{-0.35cm}
\caption{ The electrophoretic mobility of the rod-shaped macroion is shown
against ionic strength of monovalent salt $ n_{sI} $. The macroion is an 
{\it infinite} rod with the radius $ R_{rod}=5a $. The surface charge density 
of the macroion
is $ \sigma_{rod} \sim 0.08 e/a^{2} $ (filled circles), $ 0.06e/a^{2} $ (open 
circles), and $ 0.04 e/a^{2} $ (filled squares), which correspond to 
0.66C/m$ {}^{2} $, 0.49C/m$ {}^{2} $ and 0.33C/m$ {}^{2} $, respectively.
The normalization of the mobility is $ \mu_{0} \approx 21 (\mu{\rm m/sec})
/({\rm V/cm}) $.}
\label{Fig.mu_Rsalt}
\end{figure}

\begin{figure}
\centerline{\scalebox{0.65}{\includegraphics{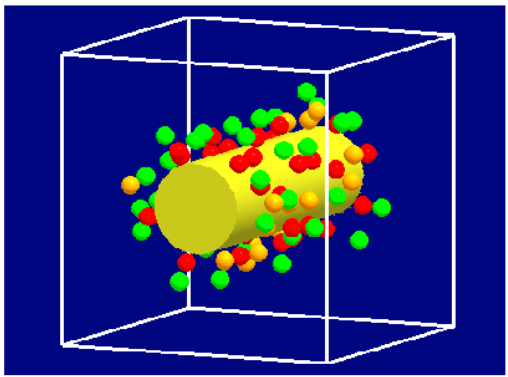}}}

\vspace*{-0.35cm}
\caption{ The bird's-eye view of the cylindrical macroion and other ions 
located within 4a from the macroion surface are shown for the run in 
Fig.\ref{Fig.mu_Rsalt}.  The surface charge density is $ \sigma= 0.08e/a^{2} $
(0.66 C/m$ {}^{2} $) and the salt ionic strength is $ n_{sI} \sim 0.006/a^{3} $
(3.6 Mol/l).  
The trivalent and monovalent counterions are drawn in red and yellow, 
respectively, and the monovalent coions in green (solvent particles are not 
drawn). }
\label{Fig.bird1sal}
\end{figure}

The charges of the {\it infinite rod} macroions in Fig.\ref{Fig.mu_Rsalt} are
$ Q_{rod}= -80e $ (filled circles), $ -60e $ (open circles) and 
$ -40e $ (filled squares).  The surface charge densities of these macroions 
are $ \sigma_{rod} = |Q_{rod}|/2 \pi R_{rod}L \sim 0.08 e/a^{2} $, $ 0.06e/a^{2} $ 
and $ 0.04 e/a^{2} $, respectively, which correspond to 0.66 C/m$ {}^{2} $, 
0.49 C/m$ {}^{2} $ and 0.33 C/m$ {}^{2} $.  For reference, the surface charge 
density of the DNA is 0.19 C/m$ {}^{2} $.
The number of Z-ions is adjusted as $ N^{+3} \approx |Q_{rod}|/eZ +30 $.
The Manning parameter $ \xi_{M}= \lambda_{B} Q_{rod}/eL $ \cite{Manning}
for these rods is between 3.9 and 7.8, and counterion condensation 
is expected which, however, does not affect the shape of the rigid-rod macroion.
We see again that the reversed mobility for the cylindrical macroion is enhanced 
by the addition of small amount of monovalent salt, similarly to the strongly 
charged spherical macroion in Fig.\ref{Fig.mu_1salt}.  
The mobility for $ \sigma_{rod}=
0.08e/a^{2} $ is still reversed at $ n_{sI} \sim 0.008/a^{3} $. 
Thus, the rod-shaped macroion is more persistent to monovalent salt than the 
spherical macroion of the same radius and surface charge density. 
The mobility of the macroion with $ \sigma_{rod}= 0.04e/a^{2} $ is not reversed,
whose surface charge density is twice large compared to that of the DNA.
However, the polyelectrolyte counterions can help to overcharge the macroion
\cite{Silva,DNA-ep}.

We compare in Fig.\ref{Fig.rdf-sal} the radial distribution functions of 
the spherical and cylindrical macroions of Fig.\ref{Fig.mu_1salt} and 
Fig.\ref{Fig.mu_Rsalt} under the same condition.
The surface charge density for these macroions is $ \sigma \sim 0.08e/a^{2} $, 
radius $ R= 5a $ and salt ionic strength $ n_{sI} \sim 0.006/a^{3} $.
The peaks of the counterion charge density are of the same height for these
cases, but the marked difference is that the monovalent counterions and coions 
are more closely concentrated to the cylindrical macroion than to the spherical 
macroion.  Consequently, for the latter with non-reversed 
mobility, two dips below the neutrality occur in the trailing tail of the 
integrated charge profile in Fig.\ref{Fig.rdf-sal}(a).  
For the former with reversed mobility, there is only one and well-defined peak 
in the integrated charge profile of Fig.\ref{Fig.rdf-sal}(b), which reveals 
strong binding of ions to the cylindrical macroion.   
This case is associated with a well-developed lateral network of Z-ions on 
the surface of the macroion, as shown in the bird's-eye view 
Fig.\ref{Fig.bird1sal}.
Majority (70\%) of the Z-ions are adsorbed on this rod macroion surface, 
which is more than sufficient for charge neutralization.  Also a small fraction 
(20\%) of the monovalent counterions are adsorbed to the macroion, unlike the 
case of the spherical macroion of the same condition.  
Thus, the network is better packed by
the counterions, which makes the surface ions less susceptible to thermal 
fluctuations and desorption.
This supports more persistence of the rod-shaped macroion to monovalent
salt than the spherical macroion of the same radius and surface charge 
density.

\section{Summary}
\label{Sec.5}

In this paper, the charge inversion phenomenon was studied by electrophoresis, 
with the focuses on the effect of the asymmetric salt in size and valence, and 
the threshold of surface charge density of the macroion for both the spherical 
and cylindrical macroions.  The criteria for the charge inversion obtained by 
the electrophoretic mobility and the radial distribution functions of ions 
coincided except around the charge inversion threshold.

First, large coions compared to the counterions played a positive role in 
enhancing the charge inversion, while the coions with larger valence than 
the counterions did a negative role.  More specifically, the reversed 
mobility started at nearly null for very small coion radius, increased linearly
with the ratio of the coion to counterion radii up to $ a^{-}/a^{+} \approx 1.5 $.  
Beyond this value, the mobility decreased sharply due to large volume fraction of
the coions and destruction of the complex by collisions.
On the other hand, the mobility was reversed and largest for the monovalent
coions, i.e. for the largest asymmetry of valences.
The reversed mobility decreased with the ratio of the coion 
to counterion valences $ Z^{-}/Z^{+} $, and flipped to normal (non-reversed) 
for the coion valence exceeding that of the counterions. The mobility was
reversed for divalent counterions and coions. 

Secondly, the addition of monovalent salt enhanced reversed mobility for a 
strongly charged macroion at small ionic strength. This was due to global 
minimization
of the electrolyte energy, and not only due to the local adjustment of ions 
on the macroion surface. On the other hand, a large amount of monovalent salt 
screened the electrostatic interactions and suppressed charge inversion.  
For a weakly charged macroion, the enhancement regime was not detected and 
reversed mobility decreased monotonically with the increase 
in the monovalent salt concentration.
  
Thirdly, the threshold of surface charge density under the monovalent salt 
was examined.  The Z-ion (multivalent counterion) correlation energy at the
threshold was several times of the thermal energy, Eq.(\ref{eq:ze}).  
A cylindrical macroion attracted counterions and coions more closely to
its vicinity, formed a well packed network of the counterions, and was more 
persistent to the monovalent salt than the spherical macroion of the same 
radius and surface charge density.

\begin{acknowledgments}
The author is highly grateful to Prof.A.Yu.Grosberg for fruitful
suggestions and discussions. He also thanks Prof.I.Ohmine for 
encouragements.
The computation of the present study was performed with the Origin 3800 
system of the University of Minnesota Supercomputing Institute, and the 
vpp800/13 supercomputer system of the Institute for Space and 
Astronautical Science (Japan). 
\end{acknowledgments}


\end{document}